\documentclass[twocolumn, secnumarabic,  floatfix, nofootinbib, nobalancelastpage, longbibliography]{revtex4-2}

\def\TitleOfPaper{Seven Useful Questions in Density Functional Theory}
\def\preprintlinklocation{http://dft.uci.edu + PAPER REF}

\usepackage{amsmath}
\usepackage{physics}
\usepackage{amssymb}
\usepackage{float}
\usepackage[dvipsnames]{xcolor}

\definecolor{TITLECOL}{rgb}{0.05,0.25,0.85}
\definecolor{CONTENTSCOL}{rgb}{0.1,0.2,0.7}
\definecolor{URLCOL}{rgb}{0,0.52,0.83}
\definecolor{LINKCOL}{rgb}{0.05,0.5,0}
\definecolor{CITECOL}{rgb}{0.25,0,0.48}
\definecolor{SECOL}{rgb}{0.07,0.31,0.80}
\definecolor{SSECOL}{rgb}{0.26,0.19,0.75}

\newcommand{\coloredtitle}[1]{\title{\textcolor{TITLECOL}{#1}}}
\newcommand{\coloredauthor}[1]{\author{\textcolor{CITECOL}{#1}}} 
\renewcommand{\sec}[1]{\section{\textcolor{SECOL}{#1}}}
\newcommand{\ssec}[1]{\subsection{\textcolor{SSECOL}{#1}}}

\usepackage{graphicx}

\def\PublicationsLink{http://dft.uci.edu/publications.php}
\def\preprintlink{ \href{\preprintlinklocation}{\TitleOfPaper} }
\def\preprinttext{~}

\usepackage{fancyhdr}
\makeatletter
\fancypagestyle{titlepage}
{	\lhead{\textsc{\preprinttext\ ~ \href{\PublicationsLink}{~}}}
	\chead{}
	\rhead{\preprintlink}
	\lfoot{}}
\makeatother
\def\preprintlink{ 
	\href{\preprintlinklocation}
        {
~}
	}

\definecolor{Green}{rgb}{0.016,0.627,0}
\definecolor{Plum}{rgb}{0.17,0,0.45}
\definecolor{LBlue}{rgb}{0,0.34,0.45}
\definecolor{Sepia}{rgb}{0.37,0.17,0.02}
\definecolor{BurntOrange}{rgb}{0.78,0.39,0}

\usepackage{hyperref}
\hypersetup{
    breaklinks=true,
    bookmarksopen=true,
    bookmarksnumbered=true,
    colorlinks=true,
    linkcolor=LINKCOL,
    linktocpage=true,
    citecolor=CITECOL,
    filecolor=magenta,      
    urlcolor=URLCOL,
}

\usepackage[normalem]{ulem}

\newcommand\rst{\bgroup\markoverwith{\textcolor{red}{\rule[0.5ex]{2pt}{0.4pt}}}\ULon}
\newcommand\bst{\bgroup\markoverwith{\textcolor{blue}{\rule[0.5ex]{2pt}{0.4pt}}}\ULon}

\newcommand\blfootnote[1]{%
  \begingroup
  \renewcommand\thefootnote{}\footnote{#1}%
  \addtocounter{footnote}{-1}%
  \endgroup
}

\def\bea{\begin{eqnarray}}
\def\eea{\end{eqnarray}}
\def\ben{\begin{equation}}
\def\een{\end{equation}}
\def\benu{\begin{enumerate}}
\def\enu{\end{enumerate}}

\def\bei{\begin{itemize}}
\def\eei{\end{itemize}}
\def\beit{\begin{itemize}}
\def\eit{\end{itemize}}
\def\benu{\begin{enumerate}}
\def\enu{\end{enumerate}}

\def\n{n}

\def\sss{\scriptscriptstyle\rm}

\def\l{^\lambda}

\def\1var{(\bx_1...\bx\N)}

\def\half{\frac{1}{2}}

\def\bp{{\bf p}}
\def\br{{\bf r}}

\def\bx{{\bf x}}

\def\x{_{\sss X}}
\def\c{_{\sss C}}
\def\s{_{\sss S}}

\def\xc{_{\sss XC}}

\def\N{_{\sss N}}
\def\H{_{\sss H}}

\def\HF{^{\rm HF}}

\def\LDA{^{\rm LDA}}

\def\TF{^{\rm TF}}

\def\unif{^{\rm unif}}

\def\ee{_{\rm ee}}

\def\d{_{\sss D}}
\def\f{_{\sss F}}

\def\sph_int{ {\int d^3 r}}

\def\intr{\int d^3r\,}
\def\intrp{\int d^3r'\,}

\DeclareMathOperator*{\argmin}{arg\,min}

\graphicspath{{figures/}}

\begin{document}
\sf
\coloredtitle{\TitleOfPaper}
\blfootnote{This
article belongs to the themed collection: \em{Mathematical Physics and Numerical
Simulation of Many-Particle Systems}; V. Bach and L. Delle Site (eds.)}


\coloredauthor{Steven Crisostomo}
\email{crisosts@uci.edu}
\affiliation{Department of Physics and Astronomy, University of California, Irvine, CA 92697, USA}

\coloredauthor{Ryan Pederson}
\affiliation{Department of Physics and Astronomy, University of California, Irvine, CA 92697, USA}

\coloredauthor{John Kozlowski}
\affiliation{Department of Chemistry, University of California, Irvine, CA 92697, USA}

\coloredauthor{Bhupalee Kalita}
\affiliation{Department of Chemistry, University of California, Irvine, CA 92697, USA}

\coloredauthor{Antonio C. Cancio}
\email{accancio@bsu.edu}
\affiliation{Department of Physics and Astronomy, Ball State University, Muncie, IN 47304, USA}

\coloredauthor{Kiril Datchev}
\email{kdatchev@purdue.edu}
\affiliation{Department of Mathematics, Purdue University, West Lafayette, IN 47907, USA}

\coloredauthor{Adam Wasserman}
\email{awasser@purdue.edu}
\affiliation{Department of Chemistry, Purdue University, West Lafayette, IN 47907, USA}
\affiliation{Department of Physics and Astronomy, Purdue University, West Lafayette, IN 47907, USA}

\coloredauthor{Suhwan Song}
\affiliation{Department of Chemistry, Yonsei University, 50 Yonsei-ro Seodaemun-gu, Seoul 03722, Korea}
\affiliation{Department of Chemistry, University of California, Irvine, CA 92697, USA}

\coloredauthor{Kieron Burke}
\email{kieron@uci.edu}
\affiliation{Department of Chemistry, University of California, Irvine, CA 92697, USA}
\affiliation{Department of Physics and Astronomy, University of California, Irvine, CA 92697, USA}

\date{\today}

\begin{abstract}
We explore a variety of unsolved problems in density functional theory where mathematicians might prove useful.   We give the background and context of the different problems, and why progress toward resolving them would help those doing computations using density functional theory. Subjects covered include the magnitude of the kinetic energy in Hartree-Fock calculations, the shape of adiabatic connection curves, using the constrained search with input densities, densities of states, the semiclassical expansion of energies, the tightness of Lieb-Oxford bounds, and how we decide the accuracy of an approximate density.
\end{abstract}
\maketitle


\sec{Introduction}
\label{sec:intro}

Density functional theory (DFT) and its applications are an enormous subject in physical sciences~\cite{JSP16,PEE20,NPSB11,ZJSW19,HCC14}. Very likely, more than 50,000 papers each year publish the results of Kohn-Sham DFT calculations~\cite{PGB14}. Given this impact, it is surprising how few resources
are devoted to further developing the theory, relative to those in machine-learning~\cite{SHMH16,JEPH21,LCBG22,SAAD22,SCSN22,CNDF22}
or quantum computing~\cite{P18, AABM19, HHYJ20, CABP21, MLAL22}. After all, even very small improvements in our current DFT approximations can have enormous impact in applications in science and technology \cite{SRP15}.

Within the space of unexplored theory, there is therefore also tremendous room for contributions from mathematicians and mathematical physicists. In this paper, we take a sampling of problems thrown up in the development of density functional
theory, all of which were encountered by the last author (KB). This list of problems differs in nature from many of those 
currently studied in the mathematical literature, as our problems have all arisen in the course of practical
calculations in the field \cite{MBAGL02,RP11,BD14,F20,FHJN21,WARTCDT22,EST23}. Solutions to some of these will have clear immediate impact.  Others are
(currently) intellectual curiosities, but of course such curiosities often lead later to very tangible results. Anyone of a mathematical inclination that is interested in
one of these problems is invited to contact the authors.

The first section introduces basic notation and reviews well-known concepts.  The problems
are sufficiently diverse that no ordering scheme is particularly helpful.   Hence no importance
should be placed on their order.   The first two problems (convexity of adiabatic connection
curves and tight bounds on exchange energies) concern conjectured exact conditions of the
exchange-correlation energy.   If such conjectures could be proven, they would guide the construction
of future approximations.    The next two concern the semiclassical limit of quantum mechanics that
is relevant to density functional approximations of electronic systems. This semiclassical expansion
is notoriously difficult, and in several key places, it appears likely it requires significant development of mathematical tools, especially asymptotic analysis.  Our fifth question is much harder to define mathematically, as it concerns measures of the accuracy of the densities in approximate
DFT calculations, and which are most useful.  Our sixth question is also somewhat vague, but addresses
the disconnect between the constrained search formulation of DFT and all other approaches to the
quantum mechanics of many-fermion systems.    Our final question is different, as it really concerns quantum chemical and many-body approaches rather than DFT.   But it is a question motivated by DFT
studies and one which, very unusally, is more easily answered in a DFT context than in standard
quantum mechanics.

\sec{Background and Notation}
\label{sec:background}
We will consider systems of electrons and nuclei in the non-relativistic,
Born-Oppenheimer limit, in the absence of external electrical and magnetic
fields.  We use Hartree atomic units.  We are interested in both finite
systems (e.g., molecules) and extended systems (e.g., bulk materials) and everything
in between.  Table I is a list of some of our acronyms.

\begin{table}[htb]
\begin{center}
\begin{tabular}{|c|c|}
\hline
    Acronym & Definition  \\
\hline
     C & Correlation      \\
\hline
     CCSD & Coupled-cluster singles and doubles      \\
\hline
     DC & Density-corrected \\
\hline
     DFT & Density functional theory      \\
\hline
     GGA & Generalized gradient approximation      \\
\hline
     HF & Hartree-Fock      \\
\hline
     IP & Ionization potential      \\
\hline
     KS & Kohn-Sham      \\
\hline
     L(S)DA & Local (spin) density approximation   \\
\hline
     LO & Lieb-Oxford      \\
\hline
     PBE & Perdew-Burke-Ernzerhof      \\
\hline
     QMC & Quantum Monte Carlo      \\
\hline
     RPA & Random phase approximation      \\
\hline
     TF & Thomas-Fermi      \\
\hline
     WKB & Wentzel-Kramers-Brillouin      \\
\hline
     X & Exchange      \\
\hline
     XC & Exchange-correlation      \\
\hline
\end{tabular}
\caption{List of acronyms used throughout this work.}
\end{center}
\end{table}

We consider Hamiltonians of the form
\begin{equation}
    \hat{H}_{v} = \hat{T} + \hat{V}\ee + \hat{V}_{v},
\end{equation}
where
\ben
\hat{T} = -\half\sum_{j=1}^{N}\nabla_{j}^{2}
\een
 is the kinetic energy operator, 
\ben
\hat{V}\ee = \half \sum_{i\neq j}\frac{1}{\vert\br_i-\br_j\vert}
\een
is the electronic interaction, and 
\ben
\hat{V}_{v} = \sum_{j=1}^{N}v(\br_{j})
\een
is a multiplicative one-body external potential, the sum of attractions
to nuclei at various positions, with differing integer positive charges. Recall that the operators $\hat{T},\hat{V}\ee,\hat{V}_{v},\hat{H}_{v}$ act on the Hilbert space ${\bigwedge}^{N}L^{2}(\mathbb{R}^{3};\mathbb{C}^{2})$ and that $\hat{T}$ and $\hat{V}\ee$ are self-adjoint. For $v(\br)$ belonging to a large set of multiplicative potentials, which includes the potentials generated by point nuclei, $\hat{V}_{v}$ and $\hat{H}_{v}$  are also self-adjoint \cite{kato51}.

 We give formulas only for non-degenerate spin-unpolarized groundstates, for simplicity. Let 
 $\Psi_{v}$ be the exact unique $N$-particle ground-state wavefunction of $\hat H_{v}$,
so that the ground-state energy is 
\ben
E[v] = \langle \Psi_{v} \vert \hat H_{v} \vert \Psi_{v} \rangle.
\een
Traditional quantum chemical methods, such as coupled-cluster \cite{K02}, are extremely
effective at finding ground-state energies to within 1 kcal/mol (0.05 eV or 1.6 mH)
for small molecules (less than 30 atoms), but scale too poorly with size to be
affordable beyond that.   
Quantum Monte Carlo methods provide a similar role for both materials \cite{NU99} and difficult molecules \cite{FMNR01}.

Density functional theory (DFT) provides an alternative path to determining ground-state properties.
Following the Levy-Lieb constrained search \cite{L79} approach, we define
\ben
F[\n] = \min_{\Psi\to\n} \langle \Psi \vert \hat T + \hat V\ee \vert \Psi \rangle,
\label{eq:universal_functional}
\een
where the search is over all normalized antisymmetric wavefunctions yielding a density $\n(\br)$, extracted from a wavefunction via
\ben
\n(\br) = N \sum_{\sigma_{1},..,\sigma_{N}}\int d^{3}r_{2}..d^{3}r_{N} 
\vert \Psi(\br\sigma_{1},..,\br_{N}\sigma_{N}) \vert^{2},
\een
where $\br_{i}$ are real-space positions and $\sigma_{i}$ are the individual spins. Here we also define $\Psi[n]$ to be the minimizer of Eq.~\eqref{eq:universal_functional} (whose uniqueness we assume) and so separately we have the kinetic energy density functional $T[n]=\langle \Psi[n] \vert \hat{T} \vert \Psi[n] \rangle$ and electronic interaction energy density functional $V\ee[n] = \langle \Psi[n] \vert \hat{V}\ee \vert\Psi[n] \rangle$, which together sum to $F[n]$. 
The ground-state energy can then be extracted from a simple search over normalized densities:
\ben
E[v] = \min_\n \left\{ F[\n] + \int d^3r\, v(\br)\, \n(\br)\right\} \label{eq:Evmin}.
\een
This scheme is exact in principle due to the Hohenberg-Kohn theorems \cite{HK64}.  Moreover, the density of a given system can be
found from the Euler-Lagrange equation by varying densities, while keeping the normalization
fixed
\ben
\frac{\delta F[\n]}{\delta \n(\br)} + v(\br) = 0.
\label{eq:Euler-Lagrange}
\een
We consider only variations in densities which conserve particle number, so that the potential is determined only up to a constant, just as in Ref.~\cite{HK64}. Here we simply assume the existence of a functional derivative, but this is a subtle point and the set of densities and potentials for which Eq.~(\ref{eq:Euler-Lagrange}) exists has been explored at length \cite{EnglischEnglischP1Derivative84,EnglischEnglischP2Derivative84,VanLeeuwenDerivative2003,E11}.  The solution to this equation can be
fed back into the functional in Eq.~(\ref{eq:Evmin}) to yield $E$.

In practice $F[\n]$ is unknown and must be approximated. 
While the original Thomas-Fermi (TF) theory \cite{T27,F27} (and variations) approximated this object directly,
essentially all modern DFT calculations use the Kohn-Sham (KS) scheme \cite{KS65}.
Here we imagine a fictitious set of non-interacting fermions with the same density
as the original system, and so satisfying an effective non-interacting Schr\"odinger
equation:
\begin{equation}
    \Big\{ -\frac{1}{2}\nabla^{2} + v\s (\br) \Big\}\phi_{i}(\br) = \epsilon_{i}\phi_{i}(\br) \label{eq:KSeq},
\end{equation}
where the $\phi_{i}$'s are orthonormal in $L^{2}$ and the $\epsilon_{i}$'s are the lowest eigenvalues of the KS Hamiltonian.
These are the celebrated KS equations, whose potential and orbitals are called the KS potential
and KS orbitals, respectively. The KS wavefunction of the (fictitious) $N$-electron system
is typically a Slater determinant of the KS orbitals, denoted $\Phi$, whose density collapses
to the sum of the squares of the orbitals:
\ben \n(\br) = 2\sum_{i=1}^N \vert \phi_{i}(\br) \vert^2. \een

We may write the functional $F[\n]$ in terms of KS quantities:
\ben
F[\n]=T\s[\n] + U[\n] + E\xc[\n],
\label{eq:XC_energy}
\een
where $T\s$ is simply the sum of KS orbital kinetic energies and $U$ is the Hartree
electrostatic self-repulsion:
\ben \label{Hartree}
U[n] = \frac{1}{2}\int \int d^{3}r d^{3}r'\, \frac{n(\br)\, n(\br')}{\vert \br - \br' \vert}.
\een
The remainder is called the exchange-correlation (XC) energy, for which there
is no simple closed form expression.  Applying Eq.~(\ref{eq:Euler-Lagrange}) to the
KS system, we find a simple expression for the KS potential:
\ben
v\s(\br) = v(\br) + v\H(\br) + v\xc(\br),
\label{KS_potential}
\een
where 
\ben
v\H(\br) = \int d^{3}r' \frac{n(\br')}{\vert \br - \br' \vert}, 
\een
is the Hartree potential 
and $v\xc(\br)$ is the (multiplicative) XC potential, given by the functional
derivative of the XC energy:
\ben 
v\xc(\br) = \frac{\delta E\xc[n]}{\delta n(\br) }.
\een
Thus, any expression of the XC energy as functional of the density, either approximate or exact \cite{WSBW13},
leads to a closed-set of self-consistent equations for any electronic system characterized by $v(\br)$ and $N$.  If these can be
solved, they can be used to predict ground-state energies and densities for any electronic
system, avoiding the need to solve the Schr\"odinger equation directly.  Due to lack of the interaction, much larger
systems can be treated with KS-DFT than are feasible with other quantum chemical methods.

There is no closed form expression for $E\xc[\n]$ in general, but hundreds of distinct approximations
have been suggested in the literature \cite{PerdewJacobsL01,toulouseXC22}. The simplest is the local density approximation (LDA), already
suggested by Kohn and Sham \cite{KS65}, in which 
\ben
E\xc\LDA[n] = \int d^3r\, e\xc\unif(\n(\br)) \label{eq:ExcLDA}
\een
and $e\xc\unif(\n)$ is the XC energy per particle of a uniform gas, which is known very accurately \cite{PZ81,VWN80,PW91}.
Most practical approximations build on this form \cite{MHGXCreview17}.

While XC is defined simply by the difference in Eq.~(\ref{eq:XC_energy}), it is traditional to separate out
a simpler contribution, called exchange (X), which dominates in a total energy calculation.
The exchange energy is defined as:
\begin{equation}
\label{eq:ExactExchange}
    E\x = -\half \intr \intrp \frac{\vert\gamma\s(\br,\br')\vert^2}{\vert\br-\br'\vert},
\end{equation}
where $\gamma\s$ is the reduced or first-order KS density matrix,
\begin{equation}
    \gamma\s(\br,\br') = 2\sum_{i}^{\text{occ}} \phi^*_{i}(\br) \, \phi_{i}(\br') ,
\end{equation}
with the sum running over all occupied KS spin-orbitals. As these are explicit expressions
in terms of orbitals and not densities, even simple exchange is not available as an explicit
density functional, and is usually approximated.

Then correlation (C) is everything else in XC.  While correlation energies are often
a tiny fraction of total energies, they can play even a dominant role in the energy
differences that are important in chemistry and materials, such as bond energies
or reaction barrier heights \cite{MHG16}. Note that the energy of the many-electron system is not the KS energy. The KS ground-state energy is the sum of KS eigenvalues, which can be corrected to yield the interacting ground-state energy. In general it has no clear variational relationship requiring it to be an upper or lower bound of the exact energy \cite{KS65}. 

While the exact XC functional is elusive, several properties and constraints have been proven or postulated over the years \cite{WYB2012}. Such ``exact conditions'' can describe features of XC in certain limits, e.g. the uniform electron gas limit, but can also be general conditions that hold for any many-electron system, e.g. $E\xc[n] \leq 0$. Knowledge of exact conditions is extremely valuable in approximate functional design \cite{PS22,KLP23}. Striving to satisfy exact conditions has been the cornerstone of the design of some of the most celebrated DFT functionals \cite{SRP15}.

In a different vein, Lieb and Simon proved an extraordinary statement about DFT in 1973 \cite{LS73}.
In a Thomas-Fermi calculation $F[n]$ is approximated by
\ben
F\TF[\n] = T\s\LDA[\n] + U[\n] \label{eq:TFfunc}
\een
and 
\ben
 T\s\LDA[n] = A\s \int d^3r\, \n^{5/3}(\br),
\een
with $A\s=(3/10)(3\pi^{2})^{2/3} $ trivially determined by the non-interacting uniform gas. The proportionality $T\s[n] \sim \int d^{3}r\,n^{5/3}(\br)$ had been shown by Thomas and Fermi, with gradient corrections proposed by von Weizs{\"a}cker \cite{W1935}, but the expression was given a precise mathematical meaning by Lieb and Simon. They showed its relative error vanishes in a very specific limit of large-$N$.  In fact,  this limit applies
to all electronic systems (atoms, molecules, or solids) and is equivalent to
taking $\zeta\to\infty$, with 
\ben
N_\zeta = \zeta N, \quad \quad \ v_\zeta(\br) = \zeta^{4/3} v(\zeta^{1/3}\br) \label{eq:nvzeta}. 
\een
For atoms (or ions), this is equivalent to taking $N\to\infty$, while keeping $N/Z$ fixed.
In this limit, the KS orbitals become increasingly semiclassical, and the contribution
to $V\ee$ of any pair of orbitals vanishes, while the total tends to a finite value.
While this may appear an arcane mathematical result, it is in fact the key to understanding
the tremendous success of modern DFT by explaining why local  (and by extension,
semi-local) approximations Eq.~(\ref{eq:ExcLDA}) work as well as they do.   It is very likely they
are the dominant term in an asymptotic expansion around this limit.

The adiabatic connection formalism is a useful tool ~\cite{HJ74, LP75, GL76}, in Eq.~(\ref{eq:universal_functional}), 
 add a coupling constant $\lambda \geq 0$:
\begin{equation}
    F^{\lambda}[\n] = \min_{\Psi \rightarrow \n} \langle \Psi \vert \hat{T} + \lambda \, \hat{V}\ee \vert \Psi \rangle \, ,
\label{eq: universal functional lambda}
\end{equation}
and let $\Psi\l[n]$ denote the minimizing wavefunction for a given $\lambda$. 
For $\lambda = 1$, we have our real, physical system and ground-state density $n(\br)$. 
For $\lambda = 0$, we have the KS system, because we have turned the electron-electron
interaction off, but kept the same ground-state density $n(\br)$.  
By convention, if $\lambda$ is absent from the notation, then $\lambda = 1$ is assumed. We discuss this formalism in more detail in Section \ref{sec:convex}. 

Lastly, we mention a very simple model system, the Hubbard dimer.
The Hubbard model \cite{H63} usually refers to an extended system, in 1, 2, or 3 dimensions, with
varying lattices, and provides a paradigm for strongly correlated systems.
For example, it is often conjectured that the 2D model might capture the 
physics of superconductivity of copper oxide planes in high-temperature cuprate
superconductivity \cite{AI07}.  But a simple two-site version (i.e., zero-dimensional)
is a crude minimal-basis model for the H$_2$ molecule, with the hopping
parameter vanishing exponentially with the stretching of the bond.

Hubbard models (with infinite number of lattice sites) play an iconic role in the physics of strongly correlated systems, which dominates much of condensed matter theory \cite{HubbardReview22}. The Hubbard dimer, which can be solved analytically, is therefore ideal for illustrating many of the concepts of KS-DFT for an audience more used to thinking of many-body effects in traditional terms.
Here the strength of the Coulomb repulsion between
particles can be made arbitrarily large, without the system blowing itself apart.
It is regarded as a paradigm of Mott-Hubbard physics, where a system can be an insulator
despite having no band gap \cite{Mott61,Shastry92,AssarafMottTrans99,Kohno2DMott2012}.  Almost all popular DFT approximations become inaccurate
for such systems.   Recently, DFT treatments of Hubbard models have become popular to
understand both the exact behavior of KS treatment of such systems and the failures
of standard approximations \cite{TDHubb2013,FM14,CFSB15,DMF17,Carrascal2018,BK21,PD22}.

The simplified two-site Hubbard model (Hubbard dimer) Hamiltonian is
\begin{equation}
    \hat{H} = -t \sum_{\sigma} (\hat{c}_{1 \sigma}^{\dagger}\hat{c}_{2 \sigma} + \hat{c}_{2 \sigma}^{\dagger}\hat{c}_{1 \sigma} ) + U \sum_{i=1}^2 \hat{n}_{i \uparrow} \hat{n}_{i \downarrow} + \sum_{i=1}^2 v_i \hat{n}_{i} \, ,
\end{equation}
where $t > 0$ is the hopping parameter, $U \geq 0$ is the Hubbard $U$ parameter, and $n_i$ is the occupation at site $i$ with on-site potential $v_i$. We focus on the two-electron spin singlet case, $N=2$ and $S_z = 0$, and denote the on-site potential difference as $\Delta v = v_2 - v_1$ and occupation difference as $\Delta n = n_2 - n_1$.
Importantly for our purposes, the theorems of DFT apply to this simple
model, so it can be used to illustrate many subtle features of KS-DFT .
Although the exact energy can be derived analytically, the exact density function(al)
cannot be found explicitly (but very accurate parameterizations have been made) \cite{CFSB15}, demonstrating
that it is easier to solve the Schr\"odinger equation exactly than to find the exact functional \cite{Verstraete09}.

There are strict limitations to the value of such models for studying DFT.
While they can be used to illustrate properties of the exact functional for that system (or related
systems) and how it captures the physics of that model, it cannot be used to understand the
validity of approximations for real DFT.   This is because the Lieb-Simon limit is hard-wired
to having continuum Hamiltonians with infinite-dimensional Hilbert spaces. The Lieb-Simon proof does
not apply to any Hubbard model, no matter how extended it might be.
And we also note that with as few as four sites, there are cases where the Hohenberg-Kohn theorems fail \cite{PvL21}.

Lastly, again, we point out that there is a very long-running interest in the KS kinetic energy functional, $T\s[\n]$.  The KS equations, Eq.~(\ref{eq:KSeq}), provide the exact value of this for the given density, but at the computational cost of solving them.  If we had an explicit approximation to this object, that was sufficiently accurate, general, and inexpensive to evaluate, we could perform {\em orbital-free} DFT calculations (as is done in TF theory, i.e. Eq.~(\ref{eq:TFfunc}) inserted into Eq.~(\ref{eq:Euler-Lagrange})) for much larger systems than those for which we solve the KS equations. We do not just need $T\s[\n]$, but we also need its functional derivative, which appears in Eq.~(\ref{eq:Euler-Lagrange}) to determine the density.   

In several cases below, we study $T\s[\n]$ itself, as a functional of the density. Even if the dream of orbital-free DFT never becomes a practically useful reality, it makes a great training ground for understanding approximate functionals.  In such sections, we drop all mention of the interacting problem, and just study KS electrons in a one-body potential.  Any approximate $T\s[\n]$ can be combined with any approximate XC functional to generate approximate energies and densities for any electronic system, via solution of the Euler-Lagrange equation, Eq.~(\ref{eq:Euler-Lagrange}).

\sec{Must adiabatic connection curves be convex?}
\label{sec:convex}

We first begin with the adiabatic connection formalism, which has been an incredibly useful concept for rationalizing
and improving density functional approximations~\cite{PEB96}. Here we discuss one prominent feature, the
convexity of adiabatic connection curves. While the convexity aspect is interesting on its own,
perhaps more prominently is its implication on the relationship between the potential $U\c$ and kinetic 
$T\c$ contributions (see Eqs.~(\ref{eq:def Tc}-\ref{eq:Ec Uc Tc}) ) to $E\c$ in any many-electron system.
Namely, the kinetic contribution has never been found
to exceed half the magnitude of the potential contribution to the correlation energy, 
Eq.~(\ref{eq: tc and uc inequality}). This observation would be proved if the adiabatic connection curve was convex.

\begin{figure}[H]
\begin{center}
\includegraphics[width=8.5cm]{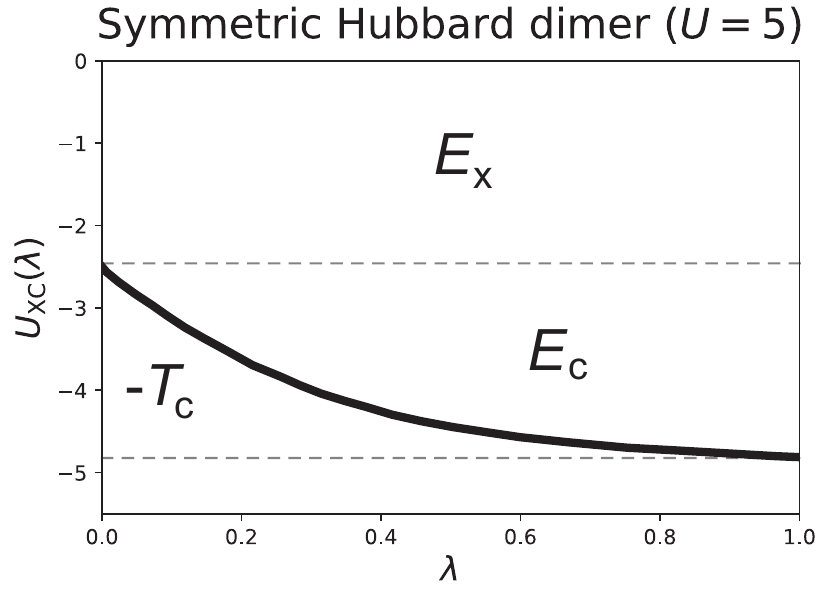}
\caption{Adiabatic connection curve (black solid line) for the symmetric ($\Delta v = 0$  and $\Delta n = 0$) Hubbard dimer with $U=5$ (obtained from Ref.~\cite{CFSB15}). Here $E\x = -U/2 = -2.5$, which is the area above the dashed line. The correlation energy $E\c$ is found from the area above the adiabatic connection curve that is below $E\x$. The kinetic contribution to the correlation energy $T\c$ is found from the area below the adiabatic connection curve that is above $U\xc(\lambda = 1)$.}
\label{fig: hubbard_dimer_U5_black_white}
\end{center}
\end{figure}

We start with the adiabatic connection formula which reads
\begin{equation}
    E\xc[n] = \int_0^1 d\lambda \, U\xc[n](\lambda) \, ,
\label{eq: adiabatic connection formula}
\end{equation}
where $U\xc[n](\lambda) = \langle \Psi\l[n] \vert \hat{V}\ee \vert \Psi\l[n] \rangle - U[n]$. Here $\Psi\l[n]$ is the ground-state wavefunction at coupling strength $\lambda$, i.e. the minimizing wavefunction of Eq.~\eqref{eq: universal functional lambda}. The integrand $U\xc[n](\lambda)$, as a function of $\lambda$, is the adiabatic connection curve mentioned above and represents the potential energy contribution to the XC energy. It is plotted exactly for the symmetric ($\Delta v = 0$  and $\Delta n = 0$) Hubbard dimer with $U=5$ in Fig.~\ref{fig: hubbard_dimer_U5_black_white}.

At $\lambda = 0$ we obtain the exchange energy, $U\xc[n](0) = E\x[n]$, and the $\lambda$-dependence is in the correlation contribution,
\begin{equation}
    U\c[n](\lambda) = U\xc[n](\lambda) - E\x[n] \, \label{eq:uxclambda} .
\end{equation}
At $\lambda = 1$, $U\xc[n] = U\xc[n](1)$ is the XC potential energy contribution of the physical system, and similarly the correlation potential energy is $U\c[n] = U\c[n](1)$. The kinetic portion of the correlation energy $T\c$ is then given by Eq.~(\ref{eq:Ec Uc Tc}). Thus, the adiabatic connection curve and integration in Eq.~\eqref{eq: adiabatic connection formula} reveal all such energy contributions to the XC energy, see Fig.~\ref{fig: hubbard_dimer_U5_black_white}. 

The adiabatic connective curve for any realistic system $U\xc[n](\lambda)$ has always been found
to be convex~\cite{LP85,FNGB05,puzder2001exchange,CFSB15,EBPb96,teale2010accurate,peach2007modeling,frydel2000adiabatic},
but this has never been proven in general.  Equivalently
\begin{equation}
    \frac{d^2 U\xc[n](\lambda)}{d\lambda ^2} \geq 0 \, \quad \text{(conjecture)} . 
\label{eq: convexity constraint}
\end{equation}
The convexity is seen clearly in the example provided in Fig.~\ref{fig: hubbard_dimer_U5_black_white}, but in realistic systems it is often more subtle~\cite{LP85,EBPb96,peach2007modeling,teale2010accurate}. The convexity also implies the following inequality
\begin{equation}
    T\c[n] \leq \frac{\vert U\c[n]\vert}{2} \, \quad \text{(conjecture)} ,
\label{eq: tc and uc inequality}
\end{equation}
which is again illustrated in Fig.~\ref{fig: hubbard_dimer_U5_black_white}.

Many popular approximate functionals, such as PBE \cite{PBE96}, violate this unproven condition~\cite{frydel2000adiabatic}. Even in the absence of a proof,
approximate functionals \textit{can} be constructed to ensure these conditions
hold and may even see improved accuracy and transferability. 
The main value in a proof (or providing counterexamples) will be in
providing physical insight on aspects of the correlation energy functional, 
and may also help clarify the relevance of such a condition. 
This awareness can be more valuable in developing approximate functionals than just
ensuring the condition is satisfied.

Searching for counterexamples of the convexity in Eq.~(\ref{eq: convexity constraint}) is challenging since obtaining high-accuracy adiabatic connection curves is demanding. Here we need to obtain high-accuracy ground-state wavefunctions with given densities, which is nontrivial, see Sec.~\ref{sec: Finding wavefunctions with given densities}. 
Insight toward proving Eq.~\eqref{eq: convexity constraint} generally may be gleaned from
a proof for the Hubbard dimer case. In Ref.~\cite{CFSB15} it was always observed that
for the general Hubbard dimer (symmetric and asymmetric versions) the adiabatic connection curve is convex.
Surely a proof is possible in such an elementary case, but we are
currently unaware of one.
Another useful direction may be in recasting the
adiabatic connection in terms of uniform coordinate scaling~\cite{LP85},
\begin{equation}
    U\xc[n](\lambda) = \frac{d}{d\lambda}\big(\lambda^2 E\xc[n_{1/\lambda}] \big) \, ,
\end{equation}
where the XC energy functional is evaluated on the uniformly scaled
density $n_{1/\lambda}(\br) = n(\br / \lambda) / \lambda^3$. 
Writing in terms of coordinate scaling instead of coupling constants is a trivial rearrangement, but can have a powerful effect on how one frames
the search for a proof.

\sec{What is the tightest Lieb-Oxford bound for exchange?}
\label{sec:LO bound}
Over the years, there has been a tremendous amount of work put into developing novel approximations to the XC energy functional, $E\xc[n]$. Although its exact form remains unknown, knowledge of its behavior in various limits has yielded immense insight into its underlying characteristics. New approximations are often formulated and constrained to satisfy exact conditions, with the aim of improving accuracy, especially for systems where data is lacking.




One constraint crucial to modern DFT has been the Lieb-Oxford (LO) lower bound~\cite{LO81}. The LO bound was originally defined for the indirect part of the total Coulomb interaction energy, and was later extended to DFT. The numerical value of the bounding constant has been slowly improved through decades of effort to construct more accurate non-empirical GGA and meta-GGA functionals. A tighter lower bound on the exchange energy has also been conjectured~\cite{PRSB14} and has been used to constrain the SCAN meta-GGA functional~\cite{SRP15}. However, the DFT community is still lacking a rigorous mathematical proof of this conjecture.  Here we give a brief overview of the problem, discuss recent advances, and review the relevant unanswered questions. 

\ssec{Lieb-Oxford bound on the potential XC energy}

The Coulomb repulsion energy for an $N$-electron wavefunction (not necessarily an antisymmetric ground state) of density $n(\br)$ can be written as a sum of the Hartree and potential XC term:
\begin{equation}
    V\ee[n] = U[n] + U\xc[n],
\end{equation}
where $U[n]$ is the classical Hartree energy defined in Eq.~(\ref{Hartree}). The original Lieb-Oxford bound~\cite{LO81} is
\begin{equation} \label{LO}
    U\xc[n] \geq - C_{\sss N} \int d^3 r \, n^{4/3} (\br),
\end{equation}
where $C_{\sss N}$ is a positive constant that depends on $N$. This expression is proportional to the local approximation to the exchange energy, given by Dirac~\cite{B29,D30}. Lieb and Oxford argued that there could not be an upper bound to $U\xc[n]$ of this form as counterexamples of wavefunctions could be easily found for which $U\xc[n]$ approached infinity, even when $n^{4/3} (\br)$ remained finite. However, the constant $C_{\sss N}$ was surprisingly difficult to calculate. Through clever derivations, $C_{\sss N}$ was shown to monotonically increase with increasing $N$, such that $C_{\sss N}\leq C_{\sss N+1}$. In Ref.~\cite{LO81}, $C_1 = 1.092$ and $C_2 \geq 1.234$. Hence $C_{\sss LO} = C_{\sss \infty} = 1.68$ defines the lower bound of $U\xc[n]$ in Eq.~(\ref{LO}). From this point onward, we will refer to the original LO constant as $C_{\sss LO}$. 
Since $E\xc \geq U\xc$, this implies~\cite{PW91}  
\begin{equation} \label{LODFT}
    E\xc[n] \geq - C_{\sss LO} \int d^3 r \, n^{4/3} (\br).
\end{equation}

This bound has been improved several times throughout the years. In the low-density limit of a uniform gas, it was found that the Coulomb repulsion is minimized by electrons forming a body-centered cubic Wigner crystal. A lower bound was found for $C_{\sss LO}$ at this limit~\cite{LP93}, giving $1.43\leq C_{\sss LO}\leq 1.68$. The upper bound was later tightened to $C_{\sss LO}\leq 1.6358$ by Chan and Handy, by replacing a trial and error estimate with  numerical evaluation~\cite{CH99}. Since then, much work has been done to shrink the range of possible $C_{\sss LO}$ values. The current possible range for this constant is $1.4442 \leq C_{\sss LO} \leq 1.5765$ \cite{LLS19, LLS22}, creating an even tighter constraint on $E\xc$. However, estimates for the upper LO bound are based on exhaustive numerical evaluations with approximate radial probability measures. It is difficult to calculate the global minimum because of the non-uniqueness of such trial states and hence the upper bound may yet shrink further~\cite{CH99, LLS22}. 

The significance of the LO bound to DFT is that it directly constrains $E\xc[\n]$ as a local functional of the electronic density, and is therefore relevant for developing advanced approximations \cite{PBE96}. As correlation is never positive, this general LO bound applies also to the exchange energy:
\begin{equation}\label{LOEX}
     E\x[n] \geq E\xc[n] \geq - C_{\sss LO} \int d^3 r \, n^{4/3} (\br).
\end{equation}

\ssec{Tighter bound on the exchange energy}
The exact exchange energy of DFT is defined as the HF exchange for the KS orbitals of a given
$\n(\br)$, as shown in Eq.~(\ref{eq:ExactExchange}). Although the current $C_{\sss LO}$ bound on $E\xc[n]$ holds for $E\x[n]$, the question of concern is whether or not this bound can be tightened in the absence of correlation. Unlike the universal LO bound, the bound on the exchange energy depends simply on spin~\cite{PS22}. $E\x[n]$ satisfies a simple spin-scaling equality that is not satisfied by $E\xc[n]$. Deriving a tighter bound for just the exchange energy could be very useful for construction of non-empirical meta-GGA density functionals~\cite{PS22}. 

The optimal exchange bound can be easily derived for an arbitrary one-electron spin-polarized system with density $n_1(\br)$, such that
\begin{equation}
    E\x[n_1] \geq -1.092\int d^3 r \, n_1^{4/3}(\br).
\end{equation}
Here, the constant term is just $C_1$ from Ref.~\cite{LO81}. Ref.~\cite{PRSB14} wrote
this bound for a two-electron spin-unpolarized ground state by using the spin-scaling relation for the exchange energy,
\begin{equation}\label{LOX_2}
\begin{split}
    E\x[n_2]  & = 2E\x[n_1] \\
    & \geq - (1.092/2^{1/3}) \int d^3 r \, n_2^{4/3}(\br) \\ 
    & = - 0.867\int d^3 r \, n_2^{4/3}(\br), \\
    \end{split}
\end{equation}
where $n_2(\br)=2 \n_1(\br)$.
This lower bound is optimally tight for compact spherical two-electron densities, as it is derived from the optimal LO bound for a one-electron system. The same work conjectured that this bound might hold for any $N$-electron, spin-unpolarized case, hence proposing a universal lower LO bound for the exchange energy,
\begin{equation}\label{LOX}
    E\x[n] \geq -0.867 \int d^3 r \, n^{4/3}(\br) \, \quad \text{(conjecture)}.
\end{equation}
We will refer to the exchange-only constant as $C_{\sss LO}^{\sss x} = 0.867$.
Since the LDA for exchange yields
\ben
E\x\LDA[\n] = -0.739 \int d^3 r \, n^{4/3}(\br),
\een
this represents a much tighter bound on the exchange energy than the one implied by the original LO Eq.~(\ref{LO}), i.e. $E\x[n]$ cannot be more than 17\% larger than its LDA counterpart.
An obvious upper bound on $E\x[n]$ is zero, but this is true only for $n = 0$. It was later shown that similar to the universal LO bound, there cannot be an upper bound on $E\x[n]$ of the form $\int d^3 r \, n^{4/3}(\br)$~\cite{PS22}. Although the conjecture for the lower bound is unproven, no contradicting examples are known~\cite{PS22}. While LDA satisfies Eq.~(\ref{LOX}), many GGA's and meta-GGA's violate it. The SCAN meta-GGA is a prominent exception, as this condition is partially built into it~\cite{SRP15}.

We next discuss the usefulness of the strictly-correlated electron limit in deriving the optimal LO bound. The determination of such bounds for $C_{\sss LO}$ and $C_{\sss LO}^{\sss x}$ is largely dependent upon initial guess states. Therefore, any method that simplifies the domain search problem for these states would be extremely useful. The original LO bound in Eq.~(\ref{LO}) is not convex in density, which can be solved by defining the Coulombic repulsion energy in terms of the Levy-Lieb functional~\cite{L83}, 
\begin{equation}
    \tilde{V}\ee[n] = \inf_{\Psi\rightarrow n}\bra{\Psi} \hat{V}_{\sss ee} \ket{\Psi}. 
\end{equation}
Here, the minimization is performed over all wavefunctions $\Psi$ yielding the density $n(\br)$. Establishing this convexity allows us to search the domain of densities (instead of the wavefunctions) for an improved lower LO bound. 

It was recently shown that $\tilde{V}\ee[n]$ yields a semiclassical limit ($\hbar\rightarrow 0$) of the Levy-Lieb functional if the domain of $\Psi$ is extended to mixed states ~\cite{L79, L83} (this semiclassical limit is very different from that of Section~\ref{sec:semiclassics}).  This relation holds for any non-negative, normalized densities of particle number $N$, with or without spin. This also corresponds to the strictly-correlated limit of DFT where the locations of the $N$ particles are completely determined by the locations of only a few ~\cite{LLS22}. In this limit, we can restrict the domain of possible states to the ground states of the classical $N$-particle problem with an external potential that is the classical equivalent of the KS potential. In fact, as it was pointed out by Seidl et al.~\cite{SBKG22}, the evaluation of $C_2$ in Ref.~\cite{LO81} was actually performed for a strictly correlated electron state with spherical density. Systematic searches could of course be performed over the strictly correlated electron densities to propose improved estimates~\cite{SVG16}.

We conclude this section with a discussion of the most recent work by Lewin et al.~\cite{LLS22}; where they propose a newer exchange-only LO bound for a general class of states with negative truncated correlations (not only Slater determinants) ~\cite{LLS22}.  With a simple proof that did not require extensive numerical evaluation, it was shown that
\begin{equation}\label{LOX_LLSa}
    E\x[n] \geq -1.249 \int d^3 r \, n^{4/3}(\br).
\end{equation}
Applying the spin-scaling relation once again, we find for spin-unpolarized systems
\begin{equation}\label{LOX_LLS}
    E\x[n] \geq -0.991 \int d^3 r \, n^{4/3}(\br).
\end{equation}
Thus $C_{\sss LO}^{\sss x}$ is no more than 0.991, which is only 34\% larger than the LDA exchange value.  

Given this information, one can imagine two possibilities. It could be that the Perdew et al.~\cite{PRSB14} conjecture is true, and that if the value in Eq.~(\ref{LOX_LLS}) is improved upon, it approaches the 2-electron value. Or, it could be that the exchange constant grows with $N$, and eventually becomes $C_{\sss LO}^{\sss x} = 0.991$ - this would mean this number is negligibly different from the optimal value. Perhaps one can find a two-orbital (4-electron) density whose exchange violates the conjecture (but not by much)? Many interesting questions are left unanswered here.

Lastly, we mention the connection with the Lieb-Simon limit of Section~\ref{sec:background}. For any given potential, as it is scaled toward the Lieb-Simon limit, the LDA exchange approximation becomes relatively more accurate. In fact, for atoms, the ratio of $E\x$ to $E\x\LDA$ monotonically decreases, approaching 1 from above as the limit is approached. This would be consistent with the conjecture that the two-electron constant is in fact a universal limit for all $N$. A study of what happens as the Lieb-Simon limit is approached  might provide further insight, or even a proof.

\sec{Can we find the semiclassical expansion of exchange-correlation?}
\label{sec:semiclassics}

From the very beginnings of DFT, one of its challenges has been
to develop a systematic methodology for
constructing functional improvements based on rigorous first principles.
Having such a tool would be a ``holy grail" of DFT, as it promises to liberate
DFT development from the curse of the \textit{ad hoc} piecing together of a patchwork of
scaling laws, constraints, and empirical data that is currently needed
to build approximations.

A promising avenue for this has been the development of
ties between DFT and semiclassical methods,~\cite{BCGP16} as we
have seen already in Sec.~\ref{sec:weyl}.
Here we explore further one example of this connection, the large-$N$ and $Z$ limit of atoms, or the so called ``statistical atom".  
The elucidation of electronic structure in this
limit has provided a justification for LDA for XC~\cite{BCGP16}, while providing insight into why GGA-level
functionals are insufficient to describe simultaneously both the energetics of atoms and molecules,
and the electronic structure of solids~\cite{PCSB06}.
Further work should provide stringent tests for new density
functionals while possibly indicating new avenues for DFT development.

The first application of DFT to a quantum
system was the treatment of neutral atoms in the limit $Z \to \infty$ using statistical
methods by Thomas and Fermi.~\cite{T27, F27}
The result was a universal scaling law $E_{\rm{TF}} \sim Z^{7/3}$, which we now know is the leading order term in an asymptotic expansion in powers of $Z^{1/3}$.
Many others have contributed since then \cite{Frank23} -- Scott~\cite{S52}, Dirac~\cite{D30},
Schwinger~\cite{S80,S81}, Englert and Schwinger~\cite{ES82,ESa84,ESb84,ESc84},
and Kunz and Rueedi~\cite{KR10}.  As a result, terms of order $Z^2$
and $Z^{5/3}$ in the total energy, and the leading order terms due to the exchange energy, also of order
$Z^{5/3}$ \cite{B29,D30},  and correlation (order $Z \log{Z}$)~\cite{KR10} have been elucidated.
Finally, the leading order term, of order $Z^{4/3}$, describing the oscillatory variation in energy across rows of the periodic table has been determined \cite{ES85}.

During this time, this expansion has been placed on a mathematically rigorous
footing.  The theorem of Lieb and Simon~\cite{LS73,LS77} discussed in Section~\ref{sec:background}, showing that the TF expression
for the total energy becomes relatively exact
under formal $\zeta$-scaling of the external potential and particle number $N$, was developed
with the statistical atom in mind. The relations of $\zeta$-scaling 
(Eq.~(\ref{eq:nvzeta})) produce the Hamiltonian of neutral atoms ($N\!=\!Z$), while scaling radial distance to produce a scale-invariant density profile (the TF density) in the $N\to\infty$ limit.
Conlon~\cite{C83} and Fefferman and Seco~\cite{FSb94} have subsequently extended the Lieb-Simon analysis to show that the HF wavefunction and energy become relatively exact in the semiclassical limit, validating the expansion for the total energy and the exchange energy through order $Z^{5/3}$ in the large-$Z$ limit (the uniform gas itself is not an example of the semiclassical limit of finite inhomogeneous systems, and can contain subtleties, even for exchange \cite{GHL19}).

We are primarily concerned with the asymptotics of the energy of non-relativistic systems, but there are many other interesting areas of investigation that we only touch upon briefly.  One concerns the corrections to the Thomas-Fermi density for large systems.  Heilman and Lieb \cite{HL95} showed that the small distance behavior of the TF density for large atoms becomes the large distance behavior of the \textit{Bohr atom}, an atom with zero electron-electron interaction potential.
This is a system with interesting questions of its own \cite{BCGP16,ARCB22arXiv} that we will not explore further here.  Likewise the extension to relativistic systems \cite{Frank23} has been of current interest to mathematicians.
The main finding related to our focus is that the non-relativistic Thomas-Fermi limit provides the leading order contribution for both energy and density even for relativistic systems.  Relativistic corrections appear to  introduce only subdominant corrections to the energy series \cite{Merz2019}.

A key to these developments is that the TF limit
is inherently a semiclassical limit -- the approach to the TF limit is
equivalent to taking $\hbar$ to zero. This allows us to employ techniques of semiclassical physics to the problem of determining the exact nature of the large $Z$ expansion, and furthermore, of developing fundamentally justifiable new approximations to DFT.
Also key is the fact that the Lieb-Simon scaling result applies to all
potentials, not just atoms, including molecules, nanostructures, or solids.  So
insights from atomic systems are potentially important for all DFT.

Although the entire energy series is of interest from the point
of view of semiclassics, the exchange and correlation contributions are of specific value for KS-DFT.  We start with exchange for neutral atoms, 
then extend to ions, and end with a few words on correlation.

\ssec{Exchange energy of neutral atoms}

We believe the large-$Z$ expansion for the exchange energy of neutral atoms, $N=Z$, can
be expressed most generally as
\ben
    E\x = -\frac{9C_2}{11} Z^{5/3} - A\x\, Z^{4/3} - B\x\, Z\log{Z} - C\x Z + \cdots
    \label{eq:exasymptotic}
\een
where $C_2 \approx 0.269900$ is the coefficient for the $Z^{5/3}$ term in the kinetic energy.
The leading term is the only term that has been derived analytically so far.
It was first found implicitly by
Bloch and independently by Dirac, both by extending the TF statistical method to exchange~\cite{B29,D30}. This leading term, not surprisingly, is exactly
given by the LDA form
of the exchange energy. Applied to the TF density of the atom:
\ben
   \ E\x = -\frac{3}{4}\left(\frac{3}{\pi}\right)^{1/3}\int d^3r\, \big[ n{\TF}(\br) \big]^{4/3}. \label{eq:ExnTF}
\een
The later $A\x$ and $C\x$ coefficients of Eq.~(\ref{eq:exasymptotic}) simply posit the continuation of the $Z^{1/3}$ series and could be produced either by corrections to the LDA X energy dependence $~n^{4/3}$,
or by the difference between the TF and exact electronic density.

The $B\x$ term is however of special interest.  Its existence
was conjectured
by Kunz and Rueddi~\cite{KR10} due to the singular behavior of the nuclear potential
at the origin, similar to the Scott correction for the total energy.
The appropriate~\cite{HL95} semiclassical limit for strongly bound
electrons is the large $N$ limit of the Bohr atom, an atom
where electron-electron interactions are
set to zero, and the necessary correction in the density matrix yields the
logarithmic behavior.  However these authors offered no details.

Recent work \cite{ARCB22arXiv, DKGSG22} has demonstrated the utility of this term
for the contribution to beyond-LDA exchange,
defined as $E\x - E\x\LDA$.  (This is of particular interest to DFT
development since it is a contribution to the asymptotic
expansion that can be used as a test for improvements in functionals.)
Exact exchange data using the OEP for atoms up to $Z=120$ yields
an excellent match to
\ben
    E\x - E\x\LDA \sim -B\x' Z\log{Z}  - C\x'Z.
\een
where $B\x' = 0.0254$~H and $C\x' = 0.0569$~H.  The former is 
almost exactly $1/4\pi^2$, and the latter is 
nearly equal to the beyond-LDA exchange energy of Hydrogen
$E\x [N\!=\!Z\!=\!1] - E\x\LDA [N\!=\!Z\!=\!1]$.
This fit is obtained using closed-shell atoms with $Z>12$, and a fit with $Z>20$ yields nearly the same results, yet these give almost
a perfect prediction for the hydrogen atom, echoing Schwinger's argument regarding
the unreasonable accuracy of asymptotic expansions~\cite{S81}.

Secondly, recent numerical work has explored in detail the
exchange energy of the Bohr atom~\cite{ARCB22arXiv}.
Here, the electron-electron coupling
$\lambda e^2$, as in Sec.~\ref{sec:convex}, is made arbitrarily small and one considers $E\x / \lambda$ in the limit $\lambda\to 0$.
In this case, both the exact exchange energy and its LDA
approximation have been recently computed to very high numerical precision,
yielding a logarithmic leading term to beyond-LDA exchange,
with coefficient
\ben
   B^{'0}\x  = \frac{1}{3 \pi^2}
\een
This is almost exactly 4/3 times the value obtained from real atoms, indicating
a likely connection.  The core region is identical in both real atoms and Bohr
atoms, as the repulsion between electrons is negligible here, and gives rise to
the bulk of the contribution.  But Bohr atoms have a finite radius beyond which
the density is identically zero, and this abrupt cut-off also contributes an additional
1/3 to $B^{'0}\x$ in the Bohr atom case.

The following questions arise.
\begin{itemize}
    \item What are the exact values of the unknown coefficients in the asymptotic expansion for exchange, i.e., $A\x$, $B\x$ and $C\x$ of Eq.~(\ref{eq:exasymptotic})?  Of immediate use to DFT, what is the beyond-LDA component of these terms ($B\x'$ and $C\x'$)? Is it even remotely
    possible to derive these coefficients?
    \item It is of nearly equal interest to generate the asymptotic expansion of the Bohr atom to the same degree of accuracy.  For example, numerical fits indicate the need of a series of high-order terms of the form $Z^{\alpha/3} \log Z$, but this is far from certain.
\end{itemize}

Reproduction of the terms of an asymptotic expansion is a stiff 
challenge for any approximate DFT.  For exchange, any GGA will yield $Z\log Z$ terms
(see Ref.~\cite{ARCB22arXiv}), but at the cost of failing to reproduce the gradient expansion for the limit of slowly varying densities.
No approximate DFT for the kinetic energy to date matches the expansion for this quantity.  Resolution of such contradictions will likely require the generation of new functional paradigms, rather than modifying existing forms.  Knowing what exactly one is aiming for is thus useful.

\ssec{Ions}
For the case of ions we need to generalize the large-$Z$ expansion to states with
net positive charge $Q=Z-N$, as
\ben E(Q,Z) = c_0(\nu) Z^{7/3} + \frac{1}{2}Z^2 + c_2(\nu) Z^{5/3} +\cdots ,\een
where $\nu = Q/Z$. The ionization potential (IP) is defined by
\ben
 I = E(1,Z) - E(0,Z).
\een
For fixed $Q$, as $Z \to \infty$, both $\nu$ and $\mu \to 0$, so one might 
expect the IP to vanish in this limit.
In Thomas-Fermi theory, this difference tends, rather,
to a constant value as shown by Lieb~\cite{L81}, and Englert in his book~\cite{E88}.  
Extended Thomas-Fermi theory, including a correction from 
$c_2(1/Z)$, yields $I = 3.15$~eV in the limit $Z\to \infty$ ~\cite{E88}. Work of Friesecke et al. \cite{Friesecke09,Friesecke10}, in which $N$ is fixed but $Z \to \infty$, is in a very different limit from the Lieb-Simon limit discussed here.

To connect with KS theory,  
careful numerical work by Snyder et al.~\cite{CSPB10} calculated the IPs for non-relativistic atoms extended to $Z=2938$.
They found that $I = C(p) + O(Z^{-1/3})$, with $p$ indicating specific columns
of the periodic table, i.e., a finite value, but one in which shell-structure persists
indefinitely.  Chemistry persists forever as one goes to ever deeper rows in
the periodic table (although it is much weaker than in the first two rows). Moreover, an average over columns I--VIII of the periodic table agrees roughly with the extended
Thomas-Fermi prediction.  At least for Bohr atoms, the beyond-LSDA
contribution disappears relative to the LSDA as $Z\to \infty$, suggesting that 
this should happen for real atoms as well, and extending the pattern, noted for the energies of neutral atoms, that the leading-order term is given exactly by the LDA.
This, however, remains to be proven, as is the form of the beyond-LDA contribution.  

\ssec{Correlation}
The issues raised above for exchange have a natural analog for the correlation energy, extending the reach of this useful problem.
The large-$Z$ expansion for correlation is
\ben
E\c = -A\c Z\log{Z} - B\c Z + \cdots
\een
The leading term is again given by LDA, with the order $Z\log Z$, coming from the high-density limit of the RPA.~\cite{BCGP16}  This has been shown to be true for atoms and quantum dots with $N$ electrons as $N$ is taken to infinity \cite{KR10}. A general theorem applicable to the $\zeta$ scaling of an arbitrary system has not been proven but seems reasonable. Careful many-body calculations of neutral atoms have allowed for numerical fit of $B\c$~\cite{BCGP16}.   A suitable modification of the numerical real-space correlation hole model~\cite{BPW97c}, a framework that underlies the 
construction of
the PBE~\cite{PBE96} correlation model, yields this coefficient~\cite{CCKB18}.
However, we are not aware of any rigorous proof that the leading order coefficient is solely determined by the LDA nor do we know of a direct calculation of the next order.

\sec{Can the Hartree-Fock kinetic energy ever exceed the true kinetic energy?}
\label{sec:kinetic}
Because DFT focuses on the density, all quantities are defined as functionals of that density.
This is very different from regular quantum mechanics, where all quantities are implicitly considered
functionals of the one-body potential, $v(\br)$.  Typically, this makes DFT more difficult to study,
but here we consider a case where DFT is {\em easier} than traditional quantum chemistry.

Starting from Eq.~(\ref{eq:universal_functional}), applied to the non-interacting KS electrons, we see
\ben T\s[n] = \min_{\Psi \to n}\langle \Psi \vert \hat{T} \vert \Psi \rangle, \een
where the minimum is over all possible normalized wavefunctions with density $\n(\br)$.  
The minimizer will typically be the single Slater determinant of KS orbitals, $\Phi$.
We can therefore define the correlation contribution to the kinetic energy via
\ben
T\c[\n]= T[\n]-T\s[\n],
\label{eq:def Tc}
\een
which, by definition, is non-negative.    Similarly, one can define
the potential energy contribution as
\ben
U\c[\n]=V\ee[\n] - U[\n]-E\x[\n] \leq 0,
\label{eq:def Uc}
\een
and then
\ben
E\c[\n] = U\c[n] + T\c[\n].
\label{eq:Ec Uc Tc}
\een
Since $E\c \leq 0$ by construction, $\vert U\c \vert \geq T\c$.  In the limit
of weak correlation, as $\lambda \to 0$ in Eq.~(\ref{eq:uxclambda}), $E\c \to -T\c \to U\c/2$ (i.e as $\lambda \to 0$ the ratio $T\c/\vert E\c\vert \to 1$ in this limit) \cite{FTB00}.
But $T\c$ cannot be negative,
because the many-body wavefunction does {\em not} minimize the kinetic energy alone,
while the KS Slater determinant does.

Now, the definitions in quantum chemistry are slightly different from those in DFT, and 
make it much harder to show that the correlation kinetic energy is positive. In quantum chemistry,for potential $v(\br)$, 
\ben
E\c\HF[v] = E[v] - E\HF[v],
\een
where $E\HF[v] = \langle \Phi_{v}\HF \vert \hat{H}_{v} \vert \Phi_{v}\HF\rangle$ is the HF energy and $\Phi_{v}\HF$, in the absence of symmetry-breaking, is the single Slater determinant that minimizes the expectation value of $\hat{H}_{v}$ and has density $n\HF_{v}(\br)$.  Scenarios with symmetry-breaking are of great interest and must be handled carefully, as the symmetry-broken case differs considerably \cite{BLLS94,Bach22}.  

The crucial difference between the HF and KS wavefunction is that $\Phi\HF_{v}$ minimizes the
Hamiltonian over {\em any} Slater determinant, while the KS determinant $\Phi[n]$ is restricted
to a set of orbitals that are simultaneously eigenfunctions of a single multiplicative
potential, $v\s(\br)$ in Eq.~(\ref{eq:KSeq}).   Both define the exchange energy by the same expression, Eq.~(\ref{eq:ExactExchange}), 
but with a different set of orbitals. By construction
\ben
\langle \Phi\HF_{v} \vert \hat{H}_{v} \vert \Phi\HF_{v} \rangle \leq  \langle \Phi[n_{v}] \vert \hat{H}_{v} \vert \Phi[n_{v}] \rangle, 
\een
where $n_{v}$ is the exact ground-state density of $v(\br)$, so that
\ben
E\c\HF[v] \geq E\c[n_{v}],
\een
i.e., the traditional quantum-chemical correlation energy is smaller in magnitude 
(less negative) than that of DFT \cite{GPG96}.

So, we can similarly write a quantum-chemical definition of kinetic correlation energy:
\ben
\  T\c\HF[v] =  \langle \Psi_{v}\vert \hat{T} \vert \Psi_{v} \rangle - \langle \Phi\HF_{v}\vert \hat{T} \vert \Phi\HF_{v} \rangle, 
\een
and ask the question, can this quantity ever be negative?  In other words, can the HF
kinetic energy ever be larger than the true kinetic energy?   
This question has been explored in a recent paper \cite{CLB22}. Numerical evidence suggests
this does not happen, i.e., no example was found, but can one prove it never
happens?

\begin{figure}[htb!]
\begin{center}
\includegraphics[width=8.5cm]{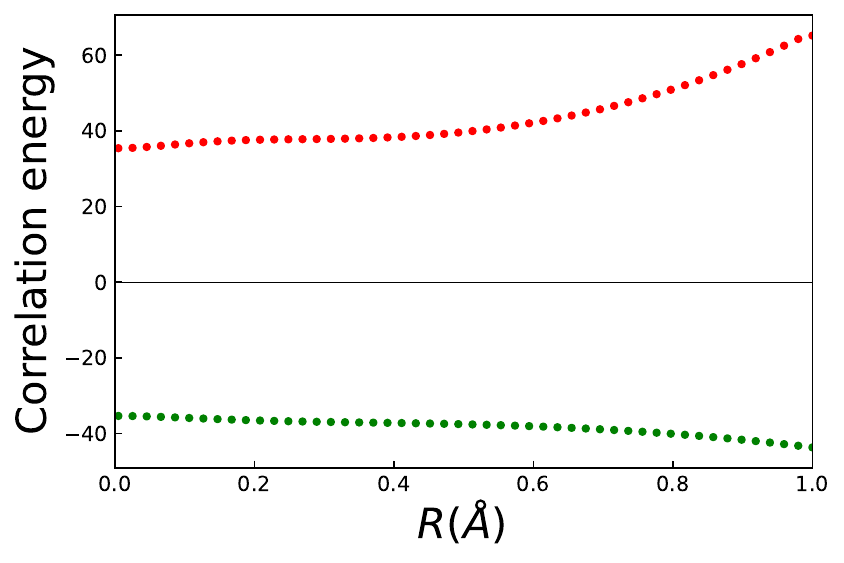}
\caption{H$_2$ molecule $E\c \HF$ (bottom) and $T\c \HF$ (top) as a function of
atomic separations, in milliHartrees. All values computed using the scheme in Ref. \cite{CLB22}.}
\end{center}
\end{figure}

Beyond intellectual curiosity, exact properties such as these are used to motivate and check
the validity of approximations. More practically, the relationship between the HF and KS kinetic energy
is immediately relevant to HF-DFT, where HF densities are used as inputs to density functionals \cite{GJPF92,SVSB22}.
In order to properly use the HF density as input, the KS orbitals yielding the HF density
should be used. Computing the KS inversion \cite{NMPS21,NSSB21,GarrigueKSInv2021,GarrigueKSInv2022,shi2022n2v} for the orbitals is generally difficult,
so usually the HF orbitals are used in place of the KS orbitals. 
Understanding the difference between $T\HF[v]$ and the $T\s[n\HF_{v}]$  becomes crucial. 
In general it appears that the difference between the two kinetic energies is small, but accuracy
is ultimately limited by the KS inversion scheme. 
A proof that $T\c\HF[v] >0$ (if it is true) would have immediate implications about the relationship between 
$T\HF[v]$ and $T\s[n_{v}]$ in general.

\sec{What makes a density accurate?}
\label{sec:accurate densities}
An important question that people running DFT calculations or developers of new DFT
approximations often raise is ``How good is my approximate density?''  It would seem to be a natural
question, given the name of the theory.   However, we will argue that there are at least as many answers
as there are askers, because the question is ill-defined without a careful choice of metric, upon
which few could agree.

Despite the name, in all the tens of thousands of (ground-state) KS-DFT calculations reported each year, 
the principal output of these calculations is the {\em energy} as a function of nuclear coordinates.
Densities are rarely reported, and if they are, they are rarely analyzed quantitatively, but rather used
to show spatial distributions that guide intuition about the underlying chemistry.

Today there are a myriad of approximate XC functionals
(more than 400), 
from basic LDA exchange to hybrid functionals \cite{LSOM18}.
This means that even if you are calculating just one single reaction,
there are more than 400 different 
approximate electron densities and energies that you can calculate.
For a small molecule, it is quite straightforward to deduce 
which XC functional gives a better reaction energy than another 
by checking the error in the DFT reaction energy from a reference reaction energy, 
often obtained from high-level electronic structure calculation methods 
or from experiment. However, for the electron density, it is not trivial to figure out
which XC functional gives an electron density closer to the exact electron density , or even what that means.
The problem with comparing an approximate density with the exact one (or a much more accurate one)
is that each density is a function of $\br$.   Thus there is an infinity of numbers to compare.

\begin{figure}[htb]
\centering
\includegraphics[width=\linewidth]{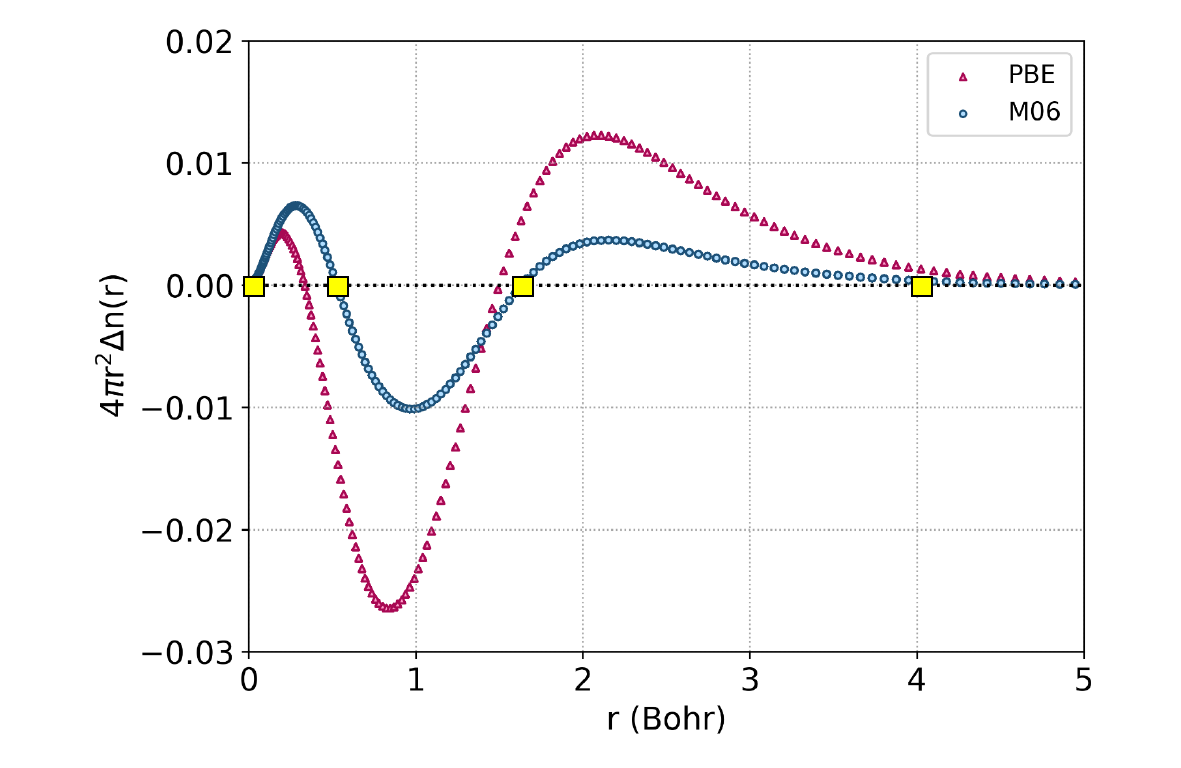}
\caption{
Radial density distribution errors in the helium atom.
CCSD with aug-cc-pVTZ basis set is used as a reference.
Radii where the M06 density crosses the CCSD density 
are marked with a square dot.
}
\label{fgr:he}
\end{figure}
In Figure~\ref{fgr:he}, the
radial density error in a helium atom is displayed
using the coupled-cluster singles and doubles (CCSD) 
electron density as the reference density.
Two different XC approximations, PBE and M06, are used
to generate approximate electron densities. 

An apparently obvious choice for comparing these two densities might be the
L2 norm \cite{MBSP17}:
\ben
\Delta_{L2}[\tilde n] = \int d^3r (\tilde n (\textbf{r})-n(\textbf{r}))^2,
\een
where $\tilde n$ denotes the approximate self-consistent electron density 
from the approximate XC functionals.  By such a measure, M06 would
clearly have a smaller ``error'' than PBE.   Moreover, if the L1 norm was used instead, M06 would still win. On the other hand, we notice that the M06 error has one more oscillation
than PBE.  So, a measure based on the derivatives of the curves might reverse the
conclusion (if not at the first derivative, then perhaps at the second, or so forth).
Thus, if we use their KS kinetic energies to judge their densities, we might
find PBE more accurate than M06.    Alternatively, we could ask which one 
yields a more accurate Hartree energy?   Even the one-body potential energy,
$\int d^3r \,  \n(\br) (-2/r)$ for the helium atom, is a different measure again.

To highlight the absurdity of the situation, we can even custom-design 
a measure of the density error to yield a certain conclusion.  Consider the
four values of $r$ at which the M06 density matches the exact density,
denoted by squares in the figure.   If we define our measure of error
as
\ben
\Delta[\tilde n] = \sum^4_{i=1} (\tilde n (r_i)-n(r_i))^2,
\een
then M06 has no error and all other approximations will be ``worse''.
Of course, we could easily choose the 3 values where the PBE density
crosses the exact density, and reverse our ranking. This argument is a little specious, as $\Delta[\tilde{n}]$ is not a distance.

Another, related issue, beyond the choice of measure, is:  Do errors in the density matter?
One can see clear differences in the figure, and one could compute various
measures of error, such as the L2 measure, but how important are such errors in practical calculations?

Having shown that there is no obvious choice of how to measure density
errors (or, more generally, density differences), the fundamental question
becomes:  Is there {\em any} choice that is well-defined and avoids arbitrariness? Indeed there is, and this can be seen by appealing to general principles of quantum
mechanics in the form of the variational principle.   For any given potential $v(\br)$,
we readily say a one trial wavefunction is ``better'' than another if its energy is lower
in the expression:

\ben
\tilde E[v] = \langle \tilde \Psi_{v} | \hat H_{v} | \tilde \Psi_{v} \rangle.
\een
Thus the energy itself provides a well-defined measure of the quality of a trial
wavefunction.   When we say one is better than another, we do not mean its nodes are more
accurate or it satisfies a cusp condition more closely, or that its asymptotic decay at large
distances is more correct.   We all understand that we simply mean its energy is
more accurate because it is lower.  Moreover, since our primary interest is in
calculating energies, our measure is quantitatively relevant:  If we expect 
an improved trial wavefunction to significantly lower the energy error on
a scale relevant to the energy we wish to extract, then the difference is important.
If, on the other hand, some improvements would reduce the energy error so little
as to be irrelevant to the problem at hand, we would regard them as unimportant.

Fortunately, this simple logic can be applied directly to KS-DFT calculations,
using the variational principle.    Consider a total energy calculated in
KS-DFT with some approximate XC.  Then the energy error is (dropping the explicit $v$-dependence for legibility)
\ben
\Delta E = \tilde E - E = \tilde E[\tilde\n] - E [\n],
\een
where the second equality simply emphasises that the KS-DFT calculation
yields an approximate energy on an approximate density, while the exact
energy is obviously the exact energy functional evaluated on the exact density.
Define the true (or functional) error of the XC approximation as

\ben
\Delta E\f[n]=\tilde E[\n]-E[\n]=\tilde E\xc[n]-E\xc[n],
\een

i.e., this is the error the functional makes when evaluated
on the exact density, i.e., removing the error in the self-consistent density.
The second equality follows from the fact that only the XC energy is
approximated in a KS calculation. Now, the difference between these two errors arises entirely from the error in the
density:
\ben
\Delta E\d[n] = \Delta E- \Delta E\f[n]=\tilde E[\tilde n]-\tilde E[n],
\label{eq:dde}
\een
providing us with a well-defined, quantitatively meaningful measure of the error
in an approximate density from a KS calculation.  This analysis has been
developed over the last decade, providing tremendous insight into many
different errors in DFT calculations, and is now called
density-corrected DFT (DC-DFT) \cite{KSB13, WNJK17,SVSB22,SSVB22}.

In practical applications in molecular calculations, it has been
found that when density-driven errors are significant for commonly-used
semilocal XC functional approximations, use of the HF-calculated density is sufficient
to greatly reduce the error. But this raises a host of questions.  For example, is the HF density leading
to some cancellation of errors, or would equally good (or possibly even better)
results be obtained with exact densities? \cite{KP23}   Another is that HF orbitals
are used in such calculations, but these are not the same as KS orbitals.
How important is the difference?    Finally, simple analytic techniques
yield much insight into what happens when XC functionals are evaluated
on densities other than self-consistent ones \cite{VSKS19}.   These must be special
cases of much more general mathematical principles and so be
applicable to many other problems where variational theorems apply.

\sec{Can we input densities into trial wavefunctions?}
\label{sec: Finding wavefunctions with given densities}

In any conference on quantum chemistry, there can be a large disconnect between those developing wavefunction approaches and those developing DFT. Essentially, whenever one approximates the XC functional in a KS calculation, one loses all sight of a wavefunction calculation.   There is no obvious way to associate the resulting DFT energy and density with some approximate wavefunction calculation.

This is odd, as the constrained search formulation is a constructive procedure for approximating $F[\n]$~\cite{L79, L83}.  Given some variational form for the wavefunction, one could calculate an approximate $F[n]$ by minimizing for a fixed density. But how does one guarantee that a given wavefunction yields a fixed density? Another deeply related question: how can an approximate $F[n]$ be generated given an appropriate approximation to $\Psi[n]$?

Both questions can be partly answered in the affirmative. The Harriman construction \cite{Harriman81} provides a prescription for generating a single Slater determinant that yields a fixed density $\n(\br)$. This assignment of densities to a subset of the Slater determinants results in an undetermined phase which affects only the kinetic energy. When the phase is extremized so that it minimizes the kinetic energy \cite{Zumbach3DHConstruction83}, the Harriman construction can be understood as an approximate mapping $\Psi[n] \approx \tilde{\Phi}[n]$. Recall that $\Psi[n]$ is defined to be the wavefunction that minimizes the expectation value of $\hat{T} +\hat{V}\ee$ and yields a fixed density $n$. Expressed slightly differently
\ben \Psi[n] = \argmin_{\Psi\to \n} \langle \Psi \vert \hat{T} + \hat{V}\ee \vert \Psi \rangle, 
\label{eq:Psin}
\een
which by definition gives $F[n] = \langle \Psi[n] \vert \hat{T} + \hat{V}\ee \vert \Psi[n] \rangle$. The Harriman construction, when properly extremized, is equivalent to Eq.~(\ref{eq:Psin}) constrained to a subset of  Slater determinants. It also defines an approximate $\tilde{F}[n] = \langle \tilde{\Phi}[n] \vert \hat{T} + \hat{V}\ee \vert \tilde{\Phi}[n] \rangle $ which is an upper bound on the exact functional, i.e. $\tilde{F}[n] \geq F[n]$ \cite{Zumbach3DHConstruction83}. While such an approximate mapping between densities and wavefunctions can be produced for a subset of Slater determinants, it does not provide a particularly accurate approximation \cite{L83,Bokanowski96}. So our first question is: can one do better? Clearly it is possible to generate approximate $F[n]$ from approximate $\Psi[n]$, but is it possible to generate standard density functional approximations in a similar manner? 

Suppose that we could generate $F\LDA[n]$ ($ = T\s\LDA[n] + U[n] + E\xc\LDA[n]$) from an approximate $\Psi[n] \approx \tilde{\Psi}[n]$, such that
\ben F\LDA[n] = \langle \tilde{\Psi}[n] \vert \hat{T} + \hat{V}\ee \vert \tilde{\Psi}[n] \rangle. \label{eq:Fapprox}\een
As an immediate consequence of the variational principle we would have the following relationship between the exact and LDA energy,
\ben E\LDA[n_{v}] = \langle \tilde{\Psi}[n_{v}] \vert \hat{H}_{v} \vert \tilde{\Psi}[n_{v}] \rangle \geq E[v] \, \quad \text{(conjecture)} \label{eq:ELDAconj}.\een
While we know of no construction for $\tilde{\Psi}[n]$, we also are unaware of any counterexample to Eq.~(\ref{eq:ELDAconj}). For atoms and ions, LDA underbinds electrons substantially, consistent with Eq.~(\ref{eq:ExnTF}) \cite{JonesGunnarsson89,NIST97}. LDA calculations on molecules typically show a reduction in this underbinding error, leading to the standard overbinding of molecules by LDA \cite{PainterAverill82,G2_97}, but not by enough to violate Eq.~(\ref{eq:ELDAconj}). If any example violating Eq.~(\ref{eq:ELDAconj}) is found, this would imply $\tilde{\Psi}[n]$ does not exist in general.

Returning to our original question, in practice QMC
or other wavefunction methods can be employed to implement the constrained search, given a constraint to yield the required density.  This could be done within e.g., variational Monte Carlo or even a neural network wavefunction.
An old trick is to calculate the ground-state wavefunction and density with one's favorite many-body method and iteratively adjust the external potential until a 
target density is achieved.  This has been an approach used by various wavefunction practitioners to make contact with DFT, by constructing XC energies via the adiabatic connection formula ~\cite{FWL90corrected,HCWR98,CS99,puzder2001exchange,teale2010accurate}, XC potentials and, via an optimized effective potential (OEP) approach, $v\x(\br)$ and $v\c(\br)$ separately \cite{FGU96}.
What is needed for the present case is to apply it to a density defined \textit{a priori} without reference to any external potential.
In fact, this has been done (albeit for one-dimensional surrogates) using the density-matrix renormalization group to solve the interacting quantum problem~\cite{WBSBW14}.
This was the first time the KS equations were solved with the exact XC functional.

But imposing a constraint after the fact is quite cumbersome, especially this one. Worse still, there can be serious convergence issues for, e.g., multireference systems~\cite{SCP03} and other strongly correlated systems~\cite{WBSBW14}, where small changes in the density
can lead to large changes in the associated KS potential. 
Even better would be if there was some automated way in which a variational wavefunction could be constructed to yield a certain density, but we know of no natural way to do this. That is to say, the density (or some set of coefficients representing the density) is an {\em input} into the procedure, not an output to be reproduced by some tedious inversion/relaxation procedure.

The basic issue is to carry out the minimization process of Eq.~(\ref{eq:universal_functional}) directly, calculating $F[n]$ from a wavefunction that yields
the correct $n(\br)$ without reference to an external potential.
A strategy to do this has been outlined by Delle Site ~\cite{DelleSite14}, following a suggestion (but not implementation) by Delle Site and David Ceperley~\cite{DGC13}. The method requires an initial wavefunction (perhaps noninteracting) that yields a target density, and then, in 
the context of QMC, a method to sample changes in the wavefunction that do not change the density.  This has been implemented for the toy case of two interacting particles in one dimension~\cite{CM16}.
This approach, unfortunately, will need a lot of work to be effective for realistic systems.

Returning to working the logic in the opposite direction, we want to be able to construct a class of  approximate wavefunctions that yield the ground-state density and energy of a given approximate density functional , as  in Eq.~(\ref{eq:Fapprox}).
Any progress constructing such a bridge would likely generate much faster 
progress in density functional development and shed light on the nature of wavefunction approximations at the same time.  Steps in this direction have been taken by Irene D'Amico \cite{DCFC11,SD16}.

A second interesting approach would be to consider the exact factorization of the wavefunction to see if
progress could be made \cite{LPS84,LM20,RG21}. If one can
express the constrained search in terms of a single integral over real space, and perform searches inside the
integral, this might be fruitful.


\sec{Are densities of particles related to densities of states?}
\label{sec:weyl}

Despite  many attempts and recent progress \cite{wesolowski2013recent,karasiev2015frank}, predictive orbital-free DFT has not yet been realized in practice in general, because sufficiently accurate approximations for $T\s[n]$ are unavailable as explicit density functionals.  Thomas-Fermi theory (TF) \cite{
T27,F27}, the first DFT, was orbital-free, and becomes relatively exact for the energies of atoms as the nuclear charge $Z\to\infty$ but it leads to atomic densities with no shell structure and, worse, TF predicts that stable molecules cannot exist \cite{Teller1962}. Extended Thomas-Fermi models \cite{brack2018semiclassical} generally improve the binding energies of molecules but remain insufficiently accurate in practice for most {\em ab-initio} quantum-chemical applications.
We now have many modern XC approximations that yield chemically useful information
in electronic structure problems, so it would be great if we could combine them
with a method that avoids solving the KS equations for the KS electrons.
An intriguing possibility for addressing this challenge is to examine the KS single-particle Hamiltonian in phase space.

\ssec{Semiclassical expansions of \texorpdfstring{$T\s[n]$}{Ts[n]}}
First, we define the exact  density of states for non-interacting
electrons:
\ben
g\s(\varepsilon) = 
\sum_i \delta(\varepsilon - \varepsilon_i)
\een
Next, 
define the {\em classical} density of states in $d$ dimensions \cite{landry2009semiclassical} as the phase-space integral
\ben 
g\s^{\rm cl}(\varepsilon)=\frac{1}{(2\pi)^d}\int d\bp \int d\br\, \delta(\varepsilon - H\s(\bp,\br))~,
\label{e:classical_KS_g}
\een 
where $H_s(\bp,\br)=\bp^2/2+v_s(\br)$ is the classical KS Hamiltonian, and we allow 1 electron per level (it is trivial to doubly occupy the levels, if desired). The average number of states below energy $\varepsilon$ is then given by: 
\ben 
{\cal{N}}\s^{\rm cl}(\varepsilon)=\int_{-\infty}^{\varepsilon} d\varepsilon'\, g\s^{\rm cl}(\varepsilon') ~,
\label{e:smooth-staircase}
\een
a smooth approximation to the KS staircase function 
\ben
{\cal{N}}_s(\varepsilon)=\sum_i \theta(\varepsilon - \varepsilon_i) = \sum_{i \colon \varepsilon_i\le \varepsilon} \varepsilon_i ~,
\label{e:KS-staircase}
\een
where the ${\varepsilon_i}$ are the eigenvalues of the KS equations. 
For $N$ non-interacting electrons, March and Plaskett \cite{MP56} demonstrated using WKB quantization that in 1D, or in 3D with spherically-symmetric potentials, the following 3-step procedure leads to the (non-interacting) Thomas-Fermi energy: (1) calculate $g_s^{\rm cl}(\varepsilon)$ through Eq.~(\ref{e:classical_KS_g}) for a given external potential $v(\br)$; (2) obtain the Fermi energy $\varepsilon_F$ from the condition $N={\cal{N}}_s^{\rm cl}(\varepsilon_F)$,
and (3) calculate the energy as
\ben
E_s^{\rm cl} = \int_{-\infty}^{\varepsilon_F} d\varepsilon\, \varepsilon\, g_s^{\rm cl}(\varepsilon)
\label{e:E_s_cl}
\een 
In the large-$N$ limit, Eq.~(\ref{e:E_s_cl}) yields the sum of the KS eigenvalues, which can be corrected to yield the interacting energy (see Eq.~(2.10) of Ref. \cite{KS65}):

\ben
E_s^{\rm cl}\approx E_s=\sum_i^N \varepsilon_i,
\een
with the fractional error vanishing as $N\to\infty$, consistent with the Lieb-Simon limit \cite{LS73} .

An efficient technique to estimate such sums when all states are bound was recently proposed by Berry and Burke \cite{berry2020exact}. Since the only input needed for evaluating $g_s^{\rm}(\varepsilon)$ in Eq.~(\ref{e:classical_KS_g}) is the density $n(\br)$ (for a given approximation to $v\xc[n](\br)$), the procedure of Eqs.~(\ref{e:classical_KS_g}-\ref{e:E_s_cl}) is a possible semiclassical route for the construction of $T_s[n]$.  A challenge here is that ${\cal{N}}_s^{\rm cl}(\varepsilon)$ is a smooth interpolation to ${\cal N}_s(\varepsilon)$ missing the sharp quantum features of ${\cal{N}}_s^{\rm }(\varepsilon)$ at the integers. This
is to be expected, because of its classical/TF nature.   One can also generate an approximate density from this procedure, since the derivative of $E\s$ with respect to the potential (at fixed $N$) is the density.  The lack of sharp quantum features in the classical staircase then translates into the lack of spatial quantum oscillations in the density.  We briefly discuss below the mathematical study of the asymptotics of ${\cal N}_s(\varepsilon)-{\cal N}_s^{\rm cl}(\varepsilon)$, where the same problem arises.

\ssec{Weyl asymptotic expansion of energy levels}

\begin{figure}[ht]
    \centering
    \includegraphics[width=8.5cm]{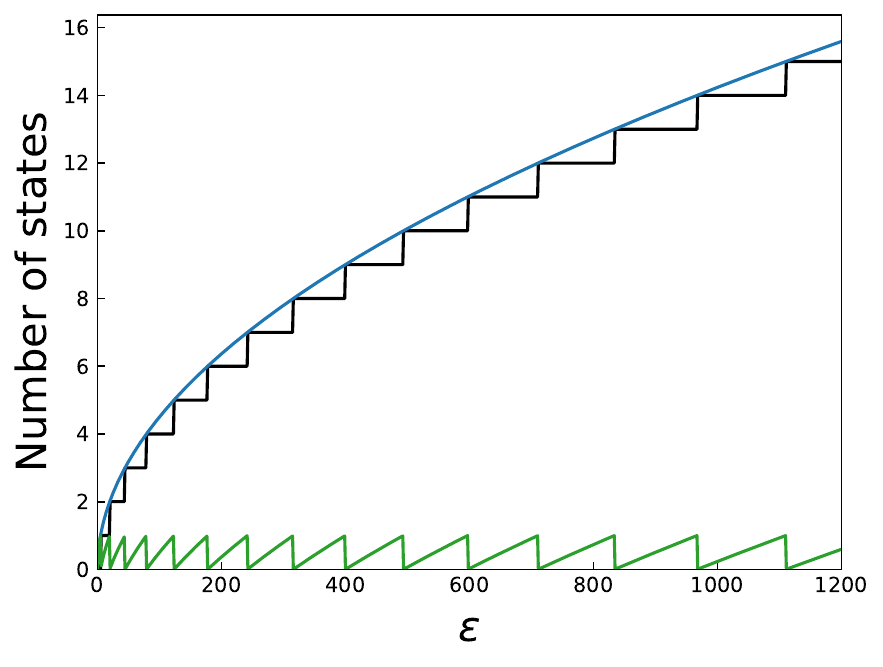}
    \caption{The black staircase function is ${\cal N}_s(\varepsilon)$, given by Eq.~\eqref{e:nsv=0}, for a one-dimensional box of length $1$, the blue parabola is ${\cal N}_s^{\rm cl}(\varepsilon)$, given by Eq.~\eqref{e:nsclv=0}, and the green zig-zag is ${\cal N}_s^{\rm cl}(\varepsilon) - {\cal N}_s(\varepsilon)$.}
    \label{fig:weyl_1d}
\end{figure}

Weyl asymptotics are asymptotics of  the staircase function ${\cal N}_s(\varepsilon)$ as $\varepsilon \to \infty$ \cite{Arendt}. 
The simplest setting is the one-dimensional Hamiltonian $-(d^2/dx^2)/2  + v(x)$ on the interval $(0,L)$ with hard walls, where $v(x)$ is an arbitrary 
potential such that $\int_0^Ldx\,\vert v(x)\vert<\infty$. Then
\begin{equation}\label{e:weyl1d}
{\cal N}_s(\varepsilon) = \frac {L\sqrt 2} \pi \sqrt \varepsilon + R(\varepsilon),
\end{equation}
where the remainder $R(\varepsilon)$ is bounded as $\varepsilon \to \infty$: see Theorem 6.1.11 of Ref.~\cite{bennewitz2020spectral}. 
Note that the leading coefficient $L\sqrt 2/ \pi$ depends on $L$ but not on $v(x)$.
%
The simplest case is $v(x)=0$, and then we get 
\begin{equation}\label{e:nsv=0}
{\cal N}_s(\varepsilon) = \sum_{i=1}^{\lfloor L\sqrt{2\varepsilon}/\pi \rfloor} 1
= \lfloor L\sqrt{2\varepsilon}/\pi \rfloor.
\end{equation}
In general, from the definition in Eq.~\eqref{e:smooth-staircase}, we find
\[
{\cal N}_s^{\rm cl}(\varepsilon) = \frac 1 \pi \int_{-\infty}^\varepsilon d \varepsilon' \int_{\{v(x)<\varepsilon'\}} dx \frac 1{\sqrt{2(\varepsilon'-v(x))}}.
\]
In the special case of $v(x)=0$, the integral is trivial, yielding the dominant term consistent with the Lieb-Simon theorem
\begin{equation}\label{e:nsclv=0}
{\cal N}^{\rm cl}_s(\varepsilon) = \frac 1 {2\pi} \int_0^L dx \int_{\{p^2 < 2\varepsilon\}} dp = \frac {L\sqrt 2}\pi \sqrt{\varepsilon};
\end{equation}
see Figure \ref{fig:weyl_1d}.

More generally, for an arbitrary but smooth $v(x)$, the classical approximation yields approximations to both the dominant term and the remainder.   Moreover, at least in one-dimension, the classical term is simply the leading contribution in the WKB semiclassical expansion \cite{Semiclassics03}, providing a full asymptotic series approximation to the
staircase, whose accuracy can be studied and controlled, using hyperasymptotic techniques.


The simplest case of a higher dimensional Weyl asymptotic is the one for the Hamiltonian $- \nabla^2/2 + v(x)$ in a two-dimensional cavity, where $v(x)$ is smooth and bounded: when $v(x)=0$ this is the problem originally studied by Weyl \cite{weyl1911}. In the typical case, one gets
\begin{equation}\label{e:weylcav}
{\cal N}_s(\varepsilon) = \frac {A}{2\pi} \varepsilon - \frac {P}{\sqrt8\pi} \sqrt{\varepsilon} + R(\varepsilon),
\end{equation}
where $A$ is the area of the cavity, $P$ is its perimeter, and $\vert R(\varepsilon)\vert /\sqrt{\varepsilon} \to 0$ as $\varepsilon \to \infty$: see equation (1.6) of Ref. \cite{Arendt} and Corollary 29.3.4 of Ref. \cite{hormander}. Compare the role played by the area and perimeter in Eq.~\eqref{e:weylcav} with that played by the length in Eq.~\eqref{e:weyl1d}, and note that again the leading coefficients depend on the cavity (through $A$ and $P$) but not on $v(x)$. See also the rest of Ref. \cite{Arendt} as well as references \cite{hormander, dimassi1999spectral, ivrii2019microlocal,frank2014cwikel} for more general results. 

\begin{figure}[ht]
    \centering
    \includegraphics[width=8.5cm]{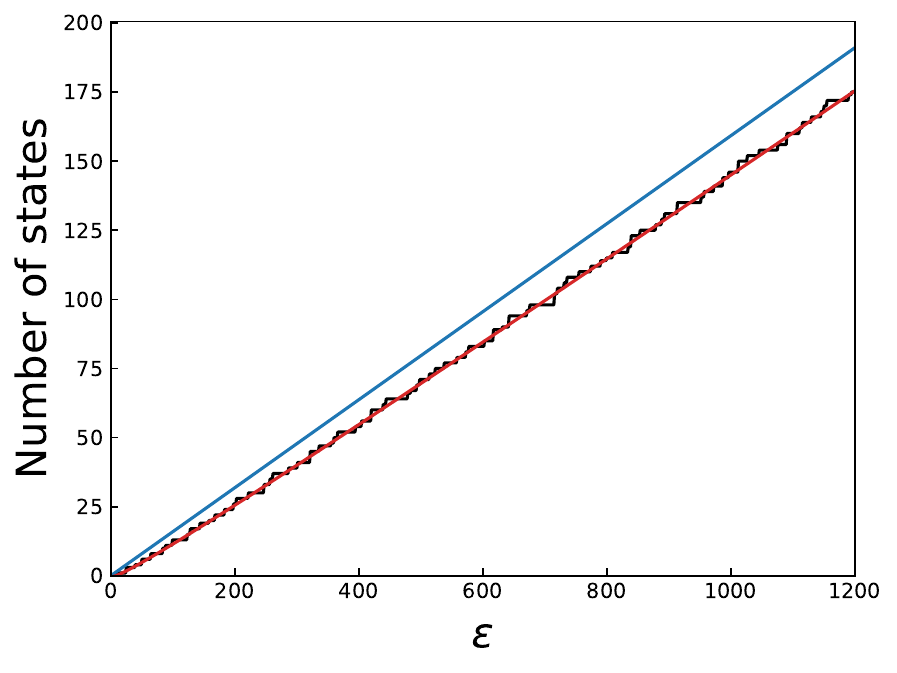} 
    \hspace{.5cm} 
    \includegraphics[width=8.5cm]{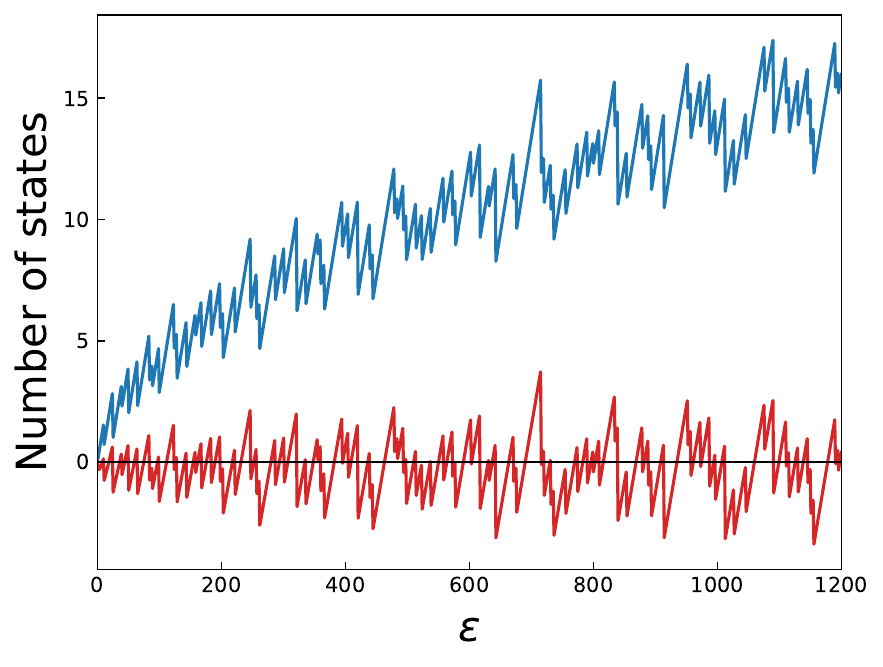}
    \caption{Above, the black staircase is ${\cal N}_s(\varepsilon)$ for a square of side length $1$, given by Eq.~\eqref{e:nssq}, the blue line is ${\cal N}_s^{\rm cl}(\varepsilon)$, given by Eq.~\eqref{e:nsclsq}, and the red curve that goes through the staircase function is ${\cal N}_s^{\rm cl}(\varepsilon) -  P\sqrt{ \varepsilon}/\sqrt{8}\pi$ (as in Eq.~\eqref{e:weylcav}). Below, the blue upper zig-zag function is ${\cal N}_s^{\rm cl}(\varepsilon) - {\cal N}_s(\varepsilon)$ and the red lower zig-zag function is  ${\cal N}_s^{\rm cl}(\varepsilon)  - P\sqrt{ \varepsilon}/\sqrt{8}\pi - {\cal N}_s(\varepsilon)$. }
    \label{fig:square}
\end{figure}

As an example, let the cavity be a square of side length $1$. Then the energy levels are  $\varepsilon_{\ell,m} = \ \pi^{2}(\ell^2 + m^2)/2$, where $\ell$ and $m$ are positive integers, and we have
\begin{equation}\label{e:nssq}
{\cal N}_s(\varepsilon) = \sum_{\ell=1}^{\lfloor \sqrt{2\varepsilon}/\pi\rfloor} \Big\lfloor  \sqrt{ \frac{2\varepsilon}{\pi^{2}} - \ell^2}\Big\rfloor,
\end{equation}
and
\begin{equation}\label{e:nsclsq}
{\cal N}_s^{\rm cl}(\varepsilon) = \frac 1 {(2\pi)^2} \int_{\{p^2 < 2\varepsilon\}} dp = \frac A{2\pi} \varepsilon;
\end{equation}
see Figure \ref{fig:square}. We see that the red is much more accurate than the blue: including the second Weyl term improves over the Thomas--Fermi approximation.

It remains an open question to understand how this improvement can be translated into an improved approximate functional for $T\s[\n]$ in three dimensions. Also, for molecular Hamiltonians, Weyl asymptotics take a different, more complicated and less detailed form \cite{Rozenblum2009}.  Molecules have open boundary conditions, and the number of classical surfaces changes as a function of bond length.  These questions are currently being worked on from many different directions \cite{OB21a, OB21b}.


\sec{Summary}
\label{sec:summary}
We hope this little tour of various issues has been useful.    These are just a handful
of issues that one could consider addressing.   They have all come from the science-side
of the subject. The important common feature is that
resolution of (almost) any of them is likely to prove useful to the field, and so 
have significant impact.   As mathematics provides an infinite variety of interesting
questions, why not focus on some that also have practical impact?
\begin{acknowledgments}
 S.C., B. K. and K. B. acknowledge support from NSF grant No. CHE-2154371. R.P. acknowledges support from DOE DE-SC0008696. J.K. acknowledges support from DOE Award No. DE-FG02-08ER46496. K.D. acknowledges support from NSF grant DMS-1708511. A.W.  acknowledges support from NSF grant CHE-1900301.

\end{acknowledgments}

\bibliography{master.bib}

\label{page:end}
\end{document}